\documentclass[a4paper,11pt]{article}

\makeatletter
\gdef\@fpheader{   }
\gdef\@journal{jhep}

\RequirePackage{amsmath}
\RequirePackage{amssymb}
\RequirePackage{epsfig}
\RequirePackage{graphicx}
\RequirePackage[numbers,sort&compress]{natbib}
\RequirePackage{color}
\RequirePackage[
colorlinks=true
,urlcolor=blue
,anchorcolor=blue
,citecolor=blue
,filecolor=blue
,linkcolor=blue
,menucolor=blue
,pagecolor=blue
,linktocpage=true
,pdfproducer=medialab
,pdfa=true
,CJKbookmarks
]{hyperref}

\newif\ifnotoc\notocfalse
\newif\ifemailadd\emailaddfalse
\newif\iftoccontinuous\toccontinuousfalse

\def\@subheader{\@empty}
\def\@keywords{\@empty}
\def\@abstract{\@empty}
\def\@xtum{\@empty}
\def\@dedicated{\@empty}
\def\@arxivnumber{\@empty}
\def\@collaboration{\@empty}
\def\@collaborationImg{\@empty}
\def\@proceeding{\@empty}
\def\@preprint{\@empty}

\newcommand{\subheader}[1]{\gdef\@subheader{#1}}
\newcommand{\keywords}[1]{\if!\@keywords!\gdef\@keywords{#1}\else%
\PackageWarningNoLine{\jname}{Keywords already defined.\MessageBreak Ignoring last definition.}\fi}
\renewcommand{\abstract}[1]{\gdef\@abstract{#1}}
\newcommand{\dedicated}[1]{\gdef\@dedicated{#1}}
\newcommand{\arxivnumber}[1]{\gdef\@arxivnumber{#1}}
\newcommand{\proceeding}[1]{\gdef\@proceeding{#1}}
\newcommand{\xtumfont}[1]{\textsc{#1}}
\newcommand{\correctionref}[3]{\gdef\@xtum{\xtumfont{#1} \href{#2}{#3}}}
\newcommand\jname{JHEP}
\newcommand\acknowledgments{\section*{Acknowledgments}}

\newcommand\preprint[1]{\gdef\@preprint{\hfill #1}}

\newcommand\note[2][]{%
\if!#1!%
\stepcounter{footnote}\footnotetext{#2}%
\else%
{\renewcommand\thefootnote{#1}%
\footnotetext{#2}}%
\fi}

\newtoks\auth@toks
\renewcommand{\author}[2][]{%
  \if!#1!%
    \auth@toks=\expandafter{\the\auth@toks#2\ }%
  \else
    \auth@toks=\expandafter{\the\auth@toks#2$^{#1}$\ }%
  \fi
}

\newtoks\affil@toks\newif\ifaffil\affilfalse
\newcommand{\affiliation}[2][]{%
\affiltrue
  \if!#1!%
    \affil@toks=\expandafter{\the\affil@toks{\item[]#2}}%
  \else
    \affil@toks=\expandafter{\the\affil@toks{\item[$^{#1}$]#2}}%
  \fi
}

\newtoks\email@toks\newcounter{email@counter}%
\newcommand{\emailAdd}[1]{%
\emailaddtrue%
\ifnum\theemail@counter>0\email@toks=\expandafter{\the\email@toks, \@email{#1}}%
\else\email@toks=\expandafter{\the\email@toks\@email{#1}}%
\fi\stepcounter{email@counter}}
\newcommand{\@email}[1]{\href{mailto:#1}{\tt #1}}

\newcommand*\collaboration[1]{\gdef\@collaboration{#1}}
\newcommand*\collaborationImg[2][]{\gdef\@collaborationImg{#2}}

\newcommand\afterLogoSpace{\smallskip}
\newcommand\afterSubheaderSpace{\vskip3pt plus 2pt minus 1pt}
\newcommand\afterProceedingsSpace{\vskip21pt plus0.4fil minus15pt}
\newcommand\afterTitleSpace{\vskip23pt plus0.06fil minus13pt}
\newcommand\afterRuleSpace{\vskip23pt plus0.06fil minus13pt}
\newcommand\afterCollaborationSpace{\vskip3pt plus 2pt minus 1pt}
\newcommand\afterCollaborationImgSpace{\vskip3pt plus 2pt minus 1pt}
\newcommand\afterAuthorSpace{\vskip5pt plus4pt minus4pt}
\newcommand\afterAffiliationSpace{\vskip3pt plus3pt}
\newcommand\afterEmailSpace{\vskip16pt plus9pt minus10pt\filbreak}
\newcommand\afterXtumSpace{\par\bigskip}
\newcommand\afterAbstractSpace{\vskip16pt plus9pt minus13pt}
\newcommand\afterKeywordsSpace{\vskip16pt plus9pt minus13pt}
\newcommand\afterArxivSpace{\vskip3pt plus0.01fil minus10pt}
\newcommand\afterDedicatedSpace{\vskip0pt plus0.01fil}
\newcommand\afterTocSpace{\bigskip\medskip}
\newcommand\afterTocRuleSpace{\bigskip\bigskip}
\newlength{\affiliationsSep}
\newcommand\beforetochook{\pagestyle{myplain}\pagenumbering{roman}}

\DeclareFixedFont\trfont{OT1}{phv}{b}{sc}{11}

\renewcommand\maketitle{
\pagestyle{empty}
\thispagestyle{titlepage}
\setcounter{page}{0}
\noindent{\small\scshape\@fpheader}\@preprint\par
\afterLogoSpace
\if!\@subheader!\else\noindent{\trfont{\@subheader}}\fi
\afterSubheaderSpace
\if!\@proceeding!\else\noindent{\sc\@proceeding}\fi
\afterProceedingsSpace
{\LARGE\flushleft\sffamily\bfseries\@title\par}
\afterTitleSpace
\hrule height 1.5\p@%
\afterRuleSpace
\if!\@collaboration!\else
{\Large\bfseries\sffamily\raggedright\@collaboration}\par
\afterCollaborationSpace
\fi
\if!\@collaborationImg!\else
{\normalsize\bfseries\sffamily\raggedright\@collaborationImg}\par
\afterCollaborationImgSpace
\fi
{\bfseries\raggedright\sffamily\the\auth@toks\par}
\afterAuthorSpace
\ifaffil\begin{list}{}{%
\setlength{\leftmargin}{0.28cm}%
\setlength{\labelsep}{0pt}%
\setlength{\itemsep}{\affiliationsSep}%
\setlength{\topsep}{-\parskip}}
\itshape\small%
\the\affil@toks
\end{list}\fi
\afterAffiliationSpace
\ifemailadd 
\noindent\hspace{0.28cm}\begin{minipage}[l]{.9\textwidth}
\begin{flushleft}
\textit{E-mail:} \the\email@toks
\end{flushleft}
\end{minipage}
\else 
\PackageWarningNoLine{\jname}{E-mails are missing.\MessageBreak Plese use \protect\emailAdd\space macro to provide e-mails.}
\fi
\afterEmailSpace
\if!\@xtum!\else\noindent{\@xtum}\afterXtumSpace\fi
\if!\@abstract!\else\noindent{\renewcommand\baselinestretch{.9}\textsc{Abstract:}}\ \@abstract\afterAbstractSpace\fi
\if!\@keywords!\else\noindent{\textsc{Keywords:}} \@keywords\afterKeywordsSpace\fi
\if!\@arxivnumber!\else\noindent{\textsc{ArXiv ePrint:}} \href{http://arxiv.org/abs/\@arxivnumber}{\@arxivnumber}\afterArxivSpace\fi
\if!\@dedicated!\else\vbox{\small\it\raggedleft\@dedicated}\afterDedicatedSpace\fi
\ifnotoc\else
\iftoccontinuous\else\newpage\fi
\beforetochook\hrule
\tableofcontents
\afterTocSpace
\hrule
\afterTocRuleSpace
\fi
\setcounter{footnote}{0}
\pagestyle{myplain}\pagenumbering{arabic}
} 

\renewcommand{\baselinestretch}{1.1}\normalsize
\divide\@tempdima\baselineskip
\@tempcnta=\@tempdima
\skip\@mpfootins = \skip\footins
\renewcommand{\@dotsep}{10000}

\newcommand\ps@myplain{
\pagenumbering{arabic}
\renewcommand\@oddfoot{\hfill-- \thepage\ --\hfill}
\renewcommand\@oddhead{}}
\let\ps@plain=\ps@myplain

\newcommand\ps@titlepage{\renewcommand\@oddfoot{}\renewcommand\@oddhead{}}

\numberwithin{equation}{section}

\renewcommand\section{\@startsection{section}{1}{\z@}%
                                   {-3.5ex \@plus -1.3ex \@minus -.7ex}%
                                   {2.3ex \@plus.4ex \@minus .4ex}%
                                   {\normalfont\large\bfseries}}
\renewcommand\subsection{\@startsection{subsection}{2}{\z@}%
                                   {-2.3ex\@plus -1ex \@minus -.5ex}%
                                   {1.2ex \@plus .3ex \@minus .3ex}%
                                   {\normalfont\normalsize\bfseries}}
\renewcommand\subsubsection{\@startsection{subsubsection}{3}{\z@}%
                                   {-2.3ex\@plus -1ex \@minus -.5ex}%
                                   {1ex \@plus .2ex \@minus .2ex}%
                                   {\normalfont\normalsize\bfseries}}
\renewcommand\paragraph{\@startsection{paragraph}{4}{\z@}%
                                   {1.75ex \@plus1ex \@minus.2ex}%
                                   {-1em}%
                                   {\normalfont\normalsize\bfseries}}
\renewcommand\subparagraph{\@startsection{subparagraph}{5}{\parindent}%
                                   {1.75ex \@plus1ex \@minus .2ex}%
                                   {-1em}%
                                   {\normalfont\normalsize\bfseries}}

\def\fnum@figure{\textbf{\figurename\nobreakspace\thefigure}}
\def\fnum@table{\textbf{\tablename\nobreakspace\thetable}}

\long\def\@makecaption#1#2{%
  \vskip\abovecaptionskip
  \sbox\@tempboxa{\small #1. #2}%
  \ifdim \wd\@tempboxa >\hsize
    \small #1. #2\par
  \else
    \global \@minipagefalse
    \hb@xt@\hsize{\hfil\box\@tempboxa\hfil}%
  \fi
  \vskip\belowcaptionskip}

\renewenvironment{thebibliography}[1]{%
\begin{oldthebibliography}{#1}%
\small%
\raggedright%
\setlength{\itemsep}{5pt plus 0.2ex minus 0.05ex}%
}%
{%
\end{oldthebibliography}%
}

\makeatother

\usepackage[T1]{fontenc} 
\usepackage{romannum} 
\usepackage{CJK}

\title{{\boldmath Scattering state and bound state of scalar field in Schwarzschild spacetime:
Exact solution}}

\author[a,b]{Wen-Du Li\footnote{Current address: College of Physics and Materials Science, Tianjin Normal University, Tianjin 300387, P.R. China (liwendu@tjnu.edu.cn)},}
\author[a,b]{Yu-Zhu Chen,}
\author[b*]{and Wu-Sheng Dai}\note{daiwusheng@tju.edu.cn.}

\affiliation[a]{Theoretical Physics Division, Chern Institute of Mathematics, Nankai University, Tianjin, 300071, P. R. China}
\affiliation[b]{Department of Physics, Tianjin University, Tianjin 300350, P.R. China}

\abstract{The main aim of this paper is twofold. (1) Exact solutions of a scalar field
in the Schwarzschild spacetime are presented. The exact wave functions of
scattering states and bound-states are presented. Besides the exact solution,
we also provide explicit approximate expressions for bound-state eigenvalues
and scattering phase shifts. (2) By virtue of the exact solutions, we give a
direct calculation for the discontinuous jump on the horizon for massive
scalar fields, while in literature such a jump is obtained from an asymptotic
solution by an analytic extension treatment.
}


\begin{document} 
\maketitle 

\flushbottom

\bigskip
\bigskip
\bigskip
\bigskip

\section{Introduction}

A massive scalar field $\Phi$ with mass $\mu$ in the background of the
Schwarzschild spacetime,%
\begin{equation}
ds^{2}=-\left(  1-\frac{2M}{r}\right)  dt^{2}+\left(  1-\frac{2M}{r}\right)
^{-1}dr^{2}+r^{2}d\theta^{2}+r^{2}\sin^{2}\theta d\phi^{2}, \label{xianyuan}%
\end{equation}
is described by the scalar equation \cite{pike2002scattering}%
\begin{equation}
\left(  \frac{1}{\sqrt{-g}}\frac{\partial}{\partial x^{\mu}}\sqrt{-g}g^{\mu
\nu}\frac{\partial}{\partial x^{\nu}}-\mu^{2}\right)  \Phi=0. \label{KG}%
\end{equation}
In this paper, we present the exact solutions of bound states and scattering
states. For bound states, we solve the\ exact bound-state wave function and
present an exact implicit expression and an asymptotic expression of the
bound-state eigenvalue. For scattering states, using the Eddington-Finkelstein
coordinates, we solve the exact solutions of the scattering wave function;
this allows us to calculate the discontinuous jump of the wave function on the
horizon exactly.

On the horizon, there is a discontinuous jump of the wave function. The
calculation of the Hawking radiation relies on the magnitude of the jump. The
calculations given by Hawking \cite{hawking1975particle} and Damour and
Ruffini \cite{damour1976black} are based on an analytic extension of
asymptotic wave functions. Concretely, they construct an asymptotic
inner-horizon wave function by an analytic extension of an asymptotic
outer-horizon wave function. In this paper, instead of the analytic extension
treatment, starting from an exact wave function obtained under
the\ Eddington-Finkelstein coordinate rather than an approximate asymptotic
one, we calculate the discontinuous jump directly.

In black hole theory, the study of scattering plays an important role
\cite{futterman1987scattering,pike2002scattering}. There are many studies on
scattering, such as the asymptotic tail \cite{koyama2001asymptotic} and
complex angular momenta of scalar scattering \cite{andersson1994complex}. The
absorption cross section of regular black holes which have event horizons but
not singularities is discussed \cite{macedo2014absorption}. Many approximate
methods are developed, such as the phase-integral method for scalar scattering
\cite{andersson1995scattering}, the propagation of a massive vector field
\cite{rosa2012massive}, massive Dirac field scattering
\cite{jin1998scattering}, the absorption cross section for scalar scattering
\cite{kuchiev2004scattering}, massive spin-half scattering
\cite{dolan2006fermion}, the WKB approximation for massive Dirac field
scattering \cite{cho2005wkb} in the Schwarzschild spacetime, a massive scalar
scattering in the Reissner-Nordstr\"{o}m spacetime \cite{benone2014absorption}%
, massless planar scalar waves scattered by a charged nonrotating black hole
\cite{crispino2009scattering}, scattering by a deformed non-rotating black
hole \cite{pei2015scattering}, and massless scalar scattering by a Kerr black
hole \cite{glampedakis2001scattering}. Scattering of spin fields and vector
fields in curved spacetime is also studied: the analogue of the Mott formula
for scattering in a Coulomb background and in the Dirac scattering by a black
hole \cite{doran2002perturbation}, massive spin-$2$ fluctuations of
Schwarzschild and slowly rotating Kerr black holes \cite{brito2013massive},
the internal stationary state of a black hole for massless Dirac fields
\cite{ahn2008black}, the quasinormal modes of electromagnetic and
gravitational perturbations of a Schwarzschild black hole in an asymptotically
anti-de Sitter spacetime \cite{cardoso2001quasinormal}, and the quasinormal
mode frequencies for the massless Dirac field in Schwarzschild-AdS spacetime
\cite{giammatteo2005dirac}. Scattering on arbitrary dimensional black holes
and on black holes with a cosmological constant is considered
\cite{okawa2011super}. Scattering between two black holes is numerically
studied \cite{okawa2011super}. The scalar field perturbations of the
$4+1$-dimensional Schwarzschild black hole immersed in a G\"{o}del universe by
the Gimon-Hashimoto solution is described \cite{konoplya2005scalar}.
Scattering method can be used in the calculation of the Hawking radiation. A
systematic scattering method for the Hawking radiation is developed by Damour
and Ruffini \cite{damour1975quantum,damour1976black}. The Hawking radiation of
a Reissner-Nordstr\"{o}m-de Sitter black hole \cite{zhao2010hawking}, the
scalar particle Hawking radiation of a BTZ black hole \cite{he2007modified},
the charged Dirac particle Hawking radiation of the Kerr-Newman black hole
\cite{zhou2008hawking}, and the distribution for particles emitted by a black
hole \cite{sannan1988heuristic} are discussed by the Damour-Ruffini method.
The Dirac particle Hawking radiation of the Kerr black hole
\cite{li2008hawking} and of the BTZ black hole \cite{li2008dirac} are
calculated by the WKB approximation. The Hawking radiation of acoustic black
holes is discussed \cite{zhang2011hawking}. The renormalized expectation
values $\left\langle T_{ab}\right\rangle $ of the relevant energy-momentum
tensor operator of a massless scalar field in the Schwarzschild spacetime is
calculated \cite{balbinot2001vacuum}. Some exact solutions are also obtained.
The analytical solution of the Regge-Wheeler equation and the Teukolsky radial
equation is obtained in Ref. \cite{fiziev2011application}. In Ref.
\cite{vieira2016confluent}, using the truncation condition of the confluent
Heun function, the authors calculate resonant frequencies for a charged scalar
field in a dyonic black hole background, and the asymptotic form of the
scattering wave function and Hawking radiation are presented. An exact
solution of the Klein-Gordon equation in Kerr-Newman spacetime is calculated,
but in which the scattering and bound-state boundary condition are not taken
into account \cite{vieira2014exact}.

The scattering of a scalar field on the Schwarzschild spacetime can also be
dealt with by the integral equation method in which the scattering phase shift
can be given explicitly \cite{li2018scalar}. The method used in the present
paper also applies to scalar fields in the Reissner-Nordstr\"{o}m spacetime
\cite{li2021scalar}.

In section \ref{equation}, the equation of a scalar field in the Schwarzschild
spacetime and the boundary condition are given. In section \ref{Heun}, as a
key step, we convert the scalar field equation in the Schwarzschild spacetime
to a confluent Heun equation. In section \ref{boundbohanshu}, we provide an
exact solution of the bound-state wave function and an explicit asymptotic
expression for the bound-state eigenvalue; the exact solution of the
bound-state eigenvalue will be given in section \ref{exacteigenvalue}. In
section \ref{scatbohanshu}, an exact solution of the scattering wave function
is given. Moreover, we also give an explicit expression of phase shift under
the weak-field approximation. In section \ref{exacteigenvalue}, we give
an\ exact implicit expression of the bound-state eigenvalue. In section
\ref{jumpcondition}, we consider the jump condition of the wave function on
the horizon and compare our result of the discontinuous jump on the horizon
with the Hawking and Damour-Ruffini treatments. The conclusions are summarized
in section \ref{conclusion}.

\section{Scalar field in Schwarzschild spacetime \label{equation}}

\subsection{Field equation}

The Schwarzschild spacetime is spherically symmetric, so we can perform a
partial-wave expansion $\Phi\left(  x^{\mu}\right)  =\frac{1}{4\pi}\sum
_{l=0}^{\infty}\sum_{m=-l}^{l}e^{-i\omega t}Y_{lm}\left(  \theta,\phi\right)
\phi_{lm}\left(  r\right)  $. When the incident wave is a plane wave, we have
\cite{pike2002scattering}:%
\begin{equation}
\Phi\left(  x^{\mu}\right)  =\sum_{l=0}^{\infty}\left(  2l+1\right)
e^{-i\omega t}P_{l}\left(  \cos\theta\right)  \phi_{l}\left(  r\right)  ,
\end{equation}
where $\phi_{l}\left(  r\right)  $ is the radial wave function satisfying the
radial equation%
\begin{equation}
\left[  \frac{1}{r^{2}}\left(  1-\frac{2M}{r}\right)  \frac{d}{dr}r^{2}\left(
1-\frac{2M}{r}\right)  \frac{d}{dr}+\omega^{2}-\left(  1-\frac{2M}{r}\right)
\mu^{2}-\left(  1-\frac{2M}{r}\right)  \frac{l\left(  l+1\right)  }{r^{2}%
}\right]  \phi_{l}\left(  r\right)  =0,\text{\ \ }r\geq2M. \label{radialeq}%
\end{equation}

There are two singularities at $r=$ $2M$ and $r\rightarrow\infty$ in the
radial equation for $r\in\left[  2M,\infty\right)  $. $r=2M$ is the horizon of
the Schwarzschild spacetime and $r\rightarrow\infty$ is the natural boundary
of space.

\subsection{Boundary condition \label{boundarycondition}}

In the problem, we need to impose two boundary conditions at $r=2M$ and
$r\rightarrow\infty$, respectively. The boundary condition at the singular
point $r=2M$ of the Schwarzschild spacetime is%
\begin{equation}
\left\vert \phi_{l}\left(  2M\right)  \right\vert <\infty, \label{bcphi2M}%
\end{equation}
i.e., $\left\vert \phi_{l}\left(  r\right)  \right\vert $ must be finite at
the singular point \cite{choudhury2004quasinormal}.

Furthermore, the boundary condition at $r=2M$ is the asymptotic solution of
the radial equation (\ref{radialeq}) at $r\rightarrow2M$
\cite{choudhury2004quasinormal,damour1976black,nollert1993quasinormal,liu2014scattering}%
. That is, $\left.  \phi_{l}\left(  r\right)  \right\vert _{r\rightarrow
2M}=\phi_{l}^{2M}\left(  r\right)  $, where $\phi_{l}^{2M}\left(  r\right)  $
is the asymptotic solution of the radial equation at $r\rightarrow2M$. The
asymptotics of the radial equation (\ref{radialeq}) at $r\rightarrow2M$ is%
\begin{equation}
\left[  \left(  1-\frac{2M}{r}\right)  \frac{d}{dr}\left(  1-\frac{2M}%
{r}\right)  \frac{d}{dr}+\omega^{2}\right]  \phi_{l}\left(  r\right)
\overset{r\rightarrow2M}{\sim}0.
\end{equation}
The solution of this asymptotic\ equation reads%
\begin{equation}
\phi_{l}\left(  r\right)  \overset{r\rightarrow2M}{\sim}e^{\pm i\omega
r_{\ast}}, \label{asyb2M}%
\end{equation}
where $r_{\ast}=r+2M\ln\left\vert \frac{r}{2M}-1\right\vert $ is the tortoise
coordinate, i.e., $\phi_{l}^{2M}\left(  r\right)  =e^{\pm i\omega r_{\ast}}$.
This gives the boundary condition at $r=2M$. The modulus $\left\vert \phi
_{l}\left(  2M\right)  \right\vert $ is finite just as required by the
boundary condition (\ref{bcphi2M}). It should be emphasized that though the
modulus of the wave function is finite, the wave function is still singular,
because, as will be seen later, there is a jump on the phase of the wave function.

The boundary condition at $r\rightarrow\infty$ determines the solution whether
a bound state or a scattering state:%
\begin{align}
\left.  \phi_{l}\left(  r\right)  \right\vert _{r\rightarrow\infty}  &
=0,\text{ \ \ \ \ bound state,}\label{BCbs}\\
\left.  \phi_{l}\left(  r\right)  \right\vert _{r\rightarrow\infty}  &
=\phi_{l}^{\infty}\left(  r\right)  ,\text{ \ scattering state,} \label{bcss}%
\end{align}
where $\phi_{l}^{\infty}\left(  r\right)  $ is the large-distance asymptotics
of the solution of the radial equation.

Furthermore, the boundary condition at $r\rightarrow\infty$ is the asymptotic
solution of the radial equation (\ref{radialeq}) at $r\rightarrow\infty$
\cite{choudhury2004quasinormal,damour1976black,nollert1993quasinormal,liu2014scattering}%
. The asymptotics of the radial equation (\ref{radialeq}) at $r\rightarrow
\infty$ is%
\begin{equation}
\frac{1}{r^{2}}\left(  1-\frac{2M}{r}\right)  \frac{d}{dr}\left[  r^{2}\left(
1-\frac{2M}{r}\right)  \frac{d}{dr}+\eta^{2}\right]  \phi_{l}\left(  r\right)
\overset{r\rightarrow\infty}{\sim}0.
\end{equation}
The solution of this asymptotic\ equation reads%
\begin{equation}
\phi_{l}\left(  r\right)  \overset{r\rightarrow\infty}{\sim}\frac{1}{r}e^{\pm
i\eta r_{\ast}}, \label{phiinfinite}%
\end{equation}
where $\eta=\sqrt{\omega^{2}-\mu^{2}}$. That is, the large-distance
asymptotics $\phi_{l}^{\infty}\left(  r\right)  =\frac{1}{r}e^{\pm i\eta
r_{\ast}}$. This is just the boundary condition at $r\rightarrow\infty$
\cite{pike2008scattering,futterman1987scattering}.

Scattering by a Schwarzschild spacetime is essentially a kind of long-range
scattering \cite{futterman1987scattering}. Recall that for potential
scattering, the large-distance asymptotics of the solution of the radial
equation, $\phi_{l}^{\infty}\left(  r\right)  $, is the same for all
short-range potential scattering, but for long-range potential scattering
\cite{liu2014scattering,li2016scattering}, like that in our case, $\phi
_{l}^{\infty}\left(  r\right)  $ is determined by the potential and different
potentials have different asymptotic solutions
\cite{li2016exact,hod2013scattering}.

\section{Converting Scalar field equation to confluent Heun equation
\label{Heun}}

\subsection{Confluent Heun equation}

The key step in solving the scalar field equation in the Schwarzschild
spacetime is to convert the radial equation (\ref{radialeq}) to a confluent
Heun equation, also called the generalized spheroidal equation
\cite{ronveaux1995heun}.

By the variable substitution $z=r/M-1$, the radial equation (\ref{radialeq})
can be converted to a confluent Heun equation,%
\begin{equation}
\frac{d}{dz}\left(  z^{2}-1\right)  \frac{d}{dz}y\left(  z\right)  +\left[
-p^{2}\left(  z^{2}-1\right)  +2p\beta z-\lambda-\frac{m^{2}+s^{2}+2msz}%
{z^{2}-1}\right]  y\left(  z\right)  =0, \label{eqyz}%
\end{equation}
where the parameters
\begin{align}
m  &  =s=2M\sqrt{-\eta^{2}-\mu^{2}},\\
\beta &  =i\frac{M}{\eta}\left(  2\eta^{2}+\mu^{2}\right)  ,\\
p  &  =-iM\eta,\\
\lambda &  =l\left(  l+1\right)  -8\eta^{2}M^{2}-6\mu^{2}M^{2}. \label{lambda}%
\end{align}
The relation between the radial wave function $\phi_{l}\left(  r\right)  $ and
the confluent Heun function $y\left(  z\right)  $ is%
\begin{equation}
\phi_{l}\left(  r\right)  =\left.  y\left(  z\right)  \right\vert _{z=r/M-1}.
\end{equation}

Corresponding to the region outside the horizon, i.e., $r\in\left[
2M,\infty\right)  $, the range of the variable $z$ in the confluent Heun
equation is $z\in\left[  1,\infty\right)  $. The confluent Heun equation
(\ref{eqyz}) has two singular points, $z=1$ and $z\rightarrow\infty$, in the
region $z\in\left[  1,\infty\right)  $ \cite{ronveaux1995heun}. These two
singular points just correspond to the two singular points of the
Schwarzschild spacetime, $r=2M$ and $r\rightarrow\infty$.

\subsection{Boundary condition}

The boundary condition of $\phi_{l}\left(  r\right)  $ is then converted to a
boundary condition of $y\left(  z\right)  $.

The boundary condition at singular point $r=2M$, Eq. (\ref{bcphi2M}), becomes%

\begin{equation}
\left\vert y\left(  1\right)  \right\vert <\infty. \label{BCy2M}%
\end{equation}
The boundary conditions for bound states, Eq. (\ref{BCbs}), and scattering
states at $r\rightarrow\infty$, Eq. (\ref{bcss}), then become%
\begin{align}
\left.  y\left(  z\right)  \right\vert _{z\rightarrow\infty}  &  =0,\text{
\ \ \ \ bound state,}\label{BCbsy}\\
\left.  y\left(  z\right)  \right\vert _{z\rightarrow\infty}  &  =y^{\infty
}\left(  z\right)  ,\text{ \ scattering state,} \label{BCsy}%
\end{align}
where $y^{\infty}\left(  z\right)  $ is the large-distance asymptotics of the
solution of the confluent Heun equation, Eq. (\ref{eqyz}).

\section{Bound state \label{boundbohanshu}}

In this section, we solve the bound-state solution. We solve the\ exact
bound-state wave function and present an exact implicit expression and an
asymptotic expression of the bound-state eigenvalue.

\subsection{Bound-state wave function}

For bound states, let%
\begin{equation}
\eta=ik,
\end{equation}
so that for bound states $\eta=-k^{2}<0$ with $k$ a real number. Then%
\begin{align}
m  &  =s=2M\sqrt{k^{2}-\mu^{2}},\label{restriction1}\\
\beta &  =-\frac{M}{k}\left(  2k^{2}-\mu^{2}\right)  ,\label{beta}\\
p  &  =kM,\label{p}\\
\lambda &  =l\left(  l+1\right)  +8M^{2}k^{2}-6M^{2}\mu^{2}.
\label{restriction2}%
\end{align}
The bound-state boundary condition is $\phi_{l}\left(  \infty\right)  =0$, or,
$y\left(  \infty\right)  =0$, given by Eqs. (\ref{BCbs}) and (\ref{BCbsy}).

The confluent Heun equation (\ref{eqyz}) with the boundary conditions
(\ref{BCy2M}) and (\ref{BCbsy}) at $z=1$ and $z=\infty$, which corresponds to
the radial equation (\ref{radialeq}) with the boundary conditions
(\ref{bcphi2M}) and (\ref{BCbs}) at $r=2M$ and $r\rightarrow\infty$, has a
solution $\Pi\left(  p,\beta,z\right)  $ satisfying $\left\vert \Pi\left(
p,\beta,1\right)  \right\vert <\infty$ and $\left\vert \Pi\left(
p,\beta,\infty\right)  \right\vert =0$, called the radial generalized
spheroidal function of $p$-type (RGSF) \cite{ronveaux1995heun}:
\begin{align}
\Pi\left(  p,\beta,z\right)   &  =N\left(  z-1\right)  ^{\left(  m+s\right)
/2}\left(  z+1\right)  ^{\left(  m-s\right)  /2}e^{-p\left(  1+z\right)
}\nonumber\\
&  \times\operatorname*{Hc}{}^{\left(  a\right)  }\left(  p,-\beta
+m+1,m+s+1,m-s+1,\sigma;\frac{z+1}{2}\right)  ,
\end{align}
where $\operatorname*{Hc}{}^{\left(  a\right)  }\left(  p,\alpha,\gamma
,\delta,\sigma;z\right)  $ is the angular confluent Heun function, $N$ is the
normalization constant, and $\sigma=\lambda+2p\left(  -2\beta+m+s+1\right)
-m\left(  m+1\right)  $ with the restriction
\begin{equation}
m+s\geq0,\text{ }m-s\geq0,\text{ }p\geq0,\text{ }\beta\in R.
\label{restriction3}%
\end{equation}

The eigenvalue of the Heun equation (\ref{eqyz}) is denoted as $\lambda
^{\left(  r\right)  }\left(  p,\beta\right)  $, where the superscript $r$
stands for the radial generalized spheroidal function of $p$-type (RGSF). It
should be noted here that the eigenvalue $\lambda^{\left(  r\right)  }\left(
p,\beta\right)  $ is the eigenvalue of the Heun equation (\ref{eqyz}) rather
than the eigenvalue of the radial equation (\ref{radialeq}).

The restrictions (\ref{p}) and (\ref{restriction3}) require that%
\begin{equation}
k>0. \label{bsr}%
\end{equation}

The bound-state wave function can be then expressed as%
\begin{align}
\Pi\left(  k,r\right)   &  =N\left(  \frac{r}{M}-2\right)  ^{2M\sqrt{k^{2}%
-\mu^{2}}}e^{-kr}\nonumber\\
&  \times\operatorname*{Hc}{}^{\left(  a\right)  }\left(  Mk,\frac{M}%
{k}\left(  2k^{2}-\mu^{2}\right)  +2M\sqrt{k^{2}-\mu^{2}}+1,4M\sqrt{k^{2}%
-\mu^{2}}+1,1,\sigma;\frac{r}{2M}\right) \nonumber\\
&  =2Ne^{2M\sqrt{k^{2}-\mu^{2}}\ln\left(  \frac{r}{2M}-1\right)  }%
e^{-kr}\nonumber\\
&  \times\operatorname*{Hc}{}^{\left(  a\right)  }\left(  Mk,1+\frac{M}%
{k}\left(  k+\sqrt{k^{2}-\mu^{2}}\right)  ^{2},1+4M\sqrt{k^{2}-\mu^{2}%
},1,\sigma;\frac{r}{2M}\right)  , \label{BSXi}%
\end{align}
where $N$ is the normalization constant and
\begin{equation}
\sigma=l\left(  l+1\right)  +2M^{2}\left(  k^{2}+\mu^{2}\right)
+8M^{2}\left(  \sqrt{k^{2}-\mu^{2}}+\frac{k}{2}\right)  ^{2}+2M\left(
k-\sqrt{k^{2}-\mu^{2}}\right)  .
\end{equation}
An asymptotic explicit expression of the eigenvalue $-k^{2}$ will be given
later in section (\ref{Eigenvalue}); an exact implicit expression of the
eigenvalue $-k^{2}$ is given by Eq. (\ref{Wr}) in section
\ref{exacteigenvalue}. After substituting the expression of\ bound-state
eigenvalue $-k^{2}$ into the wave function (\ref{BSXi}), we arrive at the
bound-state wave function.

\subsection{Bound-state eigenvalue \label{Eigenvalue}}

For bound states, the eigenvalue of the Heun equation (\ref{eqyz}) takes the
discrete values \cite{ronveaux1995heun}:%
\begin{equation}
\lambda^{\left(  r\right)  }\left(  p,\beta\right)  =\lambda_{n}\left(
p,\beta\right)  \label{lambdan}%
\end{equation}
with $n$ an integer. Together with Eqs. (\ref{beta}), (\ref{p}), and
(\ref{restriction2}), we have%
\begin{equation}
l\left(  l+1\right)  +8M^{2}k^{2}-6M^{2}\mu^{2}=\lambda_{n}\left(
Mk,-\frac{M}{k}\left(  2k^{2}-\mu^{2}\right)  \right)  . \label{lambdaandk}%
\end{equation}
The energy eigenvalue $-k^{2}$ can be solved from Eq. (\ref{lambdaandk}). Eq.
(\ref{lambdaandk}) is an implicit expression of the eigenvalue of the radical equation.

Before solving the exact result of the eigenvalue which will be given by
analyzing the analytic property of the $S$-matrix of scattering in section
\ref{exacteigenvalue}, we now give an asymptotic expression of the eigenvalue
of bound states.

Note that substituting the eigenvalue given by Eq. (\ref{lambdaandk}) into Eq.
(\ref{BSXi}) gives the bound-state wave function and, then, the radial
generalized spheroidal function of p-type (RGSF) in the bound-state wave
function (\ref{BSXi}) reduces to the Heun polynomials \cite{ronveaux1995heun}.

\subsubsection{Eigenvalue with large $Mk$}

\textit{Eigenvalue.} For a large $p=Mk$, the eigenvalue $\lambda_{n}$ has the
following asymptotics \cite{ronveaux1995heun}:%
\begin{align}
\lambda_{n}  &  =2Mk\left[  2\chi+\frac{M}{k}\left(  2k^{2}-\mu^{2}\right)
\right]  +\left\{  -2\chi\left[  \chi+\frac{M}{k}\left(  2k^{2}-\mu
^{2}\right)  \right]  +\frac{1}{2}\left[  8M^{2}\left(  k^{2}-\mu^{2}\right)
-1\right]  \right\} \nonumber\\
&  +\frac{1}{2Mk}\left\{  -\chi\left[  \chi+\frac{M}{k}\left(  2k^{2}-\mu
^{2}\right)  \right]  ^{2}-\left(  \chi^{2}-\frac{1}{4}\right)  \left[
\chi+\frac{M}{k}\left(  2k^{2}-\mu^{2}\right)  \right]  -\frac{1}{4}%
\chi\left[  1-16M^{2}\left(  k^{2}-\mu^{2}\right)  \right]  \right\}
\nonumber\\
&  +O\left(  \frac{1}{\left(  Mk\right)  ^{2}}\right)  , \label{lambdap}%
\end{align}
where%
\begin{equation}
\chi=n+\frac{1}{2} \label{kain}%
\end{equation}
with $n$ an integer. The eigenvalue $\eta^{2}=-k^{2}$ can be solved from Eqs.
(\ref{lambdaandk}) and (\ref{lambdap}) directly.

For instance, up to the first order of $\frac{1}{Mk}$ for simplicity, by Eqs.
(\ref{lambdaandk}), (\ref{lambdap}), and (\ref{kain}), we can obtain the
eigenvalue. Solving
\begin{align}
&  l\left(  l+1\right)  +2M^{2}\mu^{2}+8M^{2}\left(  k^{2}-\mu^{2}\right)
\nonumber\\
&  =2Mk\left[  2\left(  n+\frac{1}{2}\right)  +\frac{M}{k}\left(  2k^{2}%
-\mu^{2}\right)  \right]  +\left\{  -2\left(  n+\frac{1}{2}\right)  \left[
n+\frac{1}{2}+\frac{M}{k}\left(  2k^{2}-\mu^{2}\right)  \right]  \right.
\nonumber\\
&  +\left.  \frac{1}{2}\left[  8M^{2}\left(  k^{2}-\mu^{2}\right)  -1\right]
\right\}
\end{align}
gives%
\begin{equation}
k=\frac{2n+1}{2n\left(  n+1\right)  +l\left(  l+1\right)  +1}\mu^{2}M.
\end{equation}
Then the eigenvalue reads%
\begin{equation}
\eta^{2}=-k^{2}=-\left[  \frac{2n+1}{2n\left(  n+1\right)  +l\left(
l+1\right)  +1}\right]  ^{2}\mu^{4}M^{2}. \label{BSeigenvalue}%
\end{equation}

\textit{Comparison with Coulomb potential.} It is worth comparing the above
result, a Klein-Gordon particle in the Schwarzschild spacetime, with the
solution of a Klein-Gordon particle in the Coulomb potential. The large-$k$
(large-$Mk$) case is the high-energy case, i.e., the case of large $n$. With a
large $n$, Eq. (\ref{BSeigenvalue}) becomes%
\begin{equation}
\eta^{2}\sim-\frac{M^{2}\mu^{4}}{n^{2}}.
\end{equation}
While the eigenvalue of a Klein-Gordon particle in the Coulomb potential is
$\eta_{Coulomb}^{2}=\mu^{2}\left[  1+\alpha^{2}/\left(  n+\beta\right)
^{2}\right]  ^{-1/2}-\mu^{2}$, where $\alpha$ and $\beta$ are some constants
\cite{schwabl2013advanced}. For a large $n$, $\eta_{Coulomb}^{2}\sim-1/n^{2}$.
These two results are similar to each other.

\subsubsection{Eigenvalue with small $Mk$}

\textit{Eigenvalue.} For a small $p=Mk$, Eq. (\ref{eqyz}) can be approximately
written as%
\begin{equation}
\frac{d}{dz}\left(  z^{2}-1\right)  \frac{d}{dz}y\left(  z\right)  +\left(
-\lambda-2m^{2}\frac{1}{z-1}\right)  y\left(  z\right)  =0,
\end{equation}
where the relation $m=s$ given by Eq. (\ref{restriction1}) is used. This is a
hypergeometric equation \cite{olver2010nist}. The eigenvalue of this equation
reads%
\begin{equation}
\lambda_{n}=m^{2}-\left(  2n+1\right)  m+n\left(  n+1\right)  .
\label{lambdasmallp}%
\end{equation}

With Eqs. (\ref{restriction1}) and (\ref{restriction2}), Eq.
(\ref{lambdasmallp}) becomes%
\begin{equation}
l\left(  l+1\right)  +2M^{2}\mu^{2}+8M^{2}\left(  k^{2}-\mu^{2}\right)
=\left(  -2M\sqrt{k^{2}-\mu^{2}}\right)  ^{2}-\left(  2n+1\right)  \left(
-2M\sqrt{k^{2}-\mu^{2}}\right)  +n\left(  n+1\right)  .
\end{equation}
Then we have%
\begin{align}
k  &  =-\left\{  \frac{\mu^{2}}{2}+\frac{2n+1}{8M^{2}}\left[  8n\left(
n+1\right)  -4l\left(  l+1\right)  +1-8\mu^{2}M^{2}\right]  ^{1/2}\right.
\nonumber\\
&  +\left.  \frac{3n(n+1)}{4M^{2}}-\frac{l\left(  l+1\right)  }{4M^{2}}%
+\frac{1}{8M^{2}}\right\}  ^{1/2}.
\end{align}
The eigenvalue then reads%
\begin{align}
k  &  =-\left\{  \frac{\mu^{2}}{2}+\frac{2n+1}{8M^{2}}\left[  8n\left(
n+1\right)  -4l\left(  l+1\right)  +1-8\mu^{2}M^{2}\right]  ^{1/2}\right.
\nonumber\\
&  +\left.  \frac{3n(n+1)}{4M^{2}}-\frac{l\left(  l+1\right)  }{4M^{2}}%
+\frac{1}{8M^{2}}\right\}  . \label{smallMk}%
\end{align}

\textit{Comparison with Coulomb potential. }For small $k$ (small $Mk$), which
is the low-energy case, we also compare our result with the solution of a
Klein-Gordon particle in the Coulomb potential. For the low-energy case, $n$
is small. With a small $n$, Eq. (\ref{smallMk}) becomes%
\begin{equation}
\eta^{2}\sim A-Bn,
\end{equation}
where $A$ and $B$ are some constants. While the eigenvalue of a Klein-Gordon
particle in the Coulomb potential with small $n$ becomes $\eta_{Coulomb}%
^{2}\sim A^{\prime}-B^{\prime}n$. These two results are similar to each other.

Moreover, an exact implicit expression of the eigenvalue of bound states will
be given by analyzing the analytic property of the $S$-matrix of scattering in
section \ref{exacteigenvalue}.

\section{Scattering state \label{scatbohanshu}}

For scattering states,%
\begin{equation}
p=-ic,\text{ \ }\beta=i\zeta, \label{pbeta}%
\end{equation}
where $c=M\eta$ and $\zeta=\left(  M/\eta\right)  \left(  2\eta^{2}+\mu
^{2}\right)  $.

For scattering states, similarly to bound states, we also solve the radial
equation (\ref{radialeq}) with the boundary conditions at $r=2M$ and at
$r\rightarrow\infty$, respectively, replacing the bound-state boundary
condition (\ref{BCbs}) with the scattering-state boundary condition
(\ref{bcss}) at $r\rightarrow\infty$.

\subsection{Scattering wave function: Regular solution}

The regular solution for scattering in the Schwarzschild spacetime is a
solution satisfying the boundary condition at $r=2M$. As a comparison, in
common quantum-mechanical central potential scattering the regular solution
satisfies the boundary condition at the singular point $r=0$
\cite{friedrich2015scattering,li2016exact}.

At $r=2M$, the solution of the radial equation (\ref{radialeq}) still needs to
satisfy the boundary condition (\ref{bcphi2M}), but the parameters $p$ and
$\beta$ given by Eq. (\ref{pbeta}) are now pure imaginary numbers. The
solution now is the angular generalized spheroidal function of $c$-type,
$\Pi\left(  c,\zeta,z\right)  $, which can be achieved by substituting
$\eta=ik$ into Eq. (\ref{BSXi}):%
\begin{align}
\Pi\left(  \eta,r\right)   &  =2Ne^{2iM\sqrt{\eta^{2}+\mu^{2}}\ln\left(
\frac{r}{2M}-1\right)  }e^{i\eta r}\nonumber\\
&  \times\operatorname*{Hc}{}^{\left(  a\right)  }\left(  -i\eta M,1-i\frac
{M}{\eta}\left(  \eta-\sqrt{\eta^{2}+\mu^{2}}\right)  ^{2},1+4iM\sqrt{\eta
^{2}+\mu^{2}},1,\sigma;\frac{r}{2M}\right)  , \label{regularsolution}%
\end{align}
where%
\begin{equation}
\sigma=l\left(  l+1\right)  -2M^{2}\left(  \eta^{2}-\mu^{2}\right)
-8M^{2}\left(  \sqrt{\eta^{2}+\mu^{2}}-\frac{\eta}{2}\right)  ^{2}-2iM\left(
\eta+\sqrt{\eta^{2}+\mu^{2}}\right)  .
\end{equation}
This solution is an analogue of the regular solution in quantum-mechanical
scattering theory.

It should be noted that here the bound-state wave function $\Pi\left(
k,r\right)  $ given by Eq. (\ref{BSXi}) satisfies the boundary condition at
$r=2M$, but leaves alone the boundary condition at $r\rightarrow\infty$ before
substituting the bound-state eigenvalue which is determined by the boundary
condition at $r\rightarrow\infty$ into the eigenfunction. Therefore after
performing the replacement $\eta=ik$, $\Pi\left(  \eta,r\right)  $ can serve
as the regular solution in scattering.

\subsection{Scattering wave function: Irregular solution}

The irregular solution for scattering in the Schwarzschild spacetime is the
solution satisfying the boundary condition $\left.  \phi_{l}\left(  r\right)
\right\vert _{r\rightarrow\infty}=\phi_{l}^{\infty}\left(  r\right)  $ at
$r\rightarrow\infty$. As a comparison, in quantum-mechanical central potential
scattering the irregular solution satisfies the boundary condition at
$r\rightarrow\infty$ \cite{li2016exact}.

At $r\rightarrow\infty$, the bound-state boundary condition (\ref{BCbs}) is
replaced by the scattering boundary condition (\ref{bcss}) and the parameters
$p$ and $\beta$ are taken as imaginary numbers.

\subsubsection{Scattering boundary condition: Asymptotic behavior}

Before seeking exact solutions, we first investigate the large-distance
asymptotic behavior of the scattering wave function. The asymptotic solution
of the scattering wave function will serve as boundary conditions for
scattering states \cite{liu2014scattering,li2016scattering,li2016exact}.

The solution of Eq. (\ref{radialeq}) satisfying the scattering boundary
condition at $r\rightarrow\infty$ is the radial generalized spheroidal
function (RGSF) of $c$-type, $\Pi\left(  \eta,r\right)  $
\cite{ronveaux1995heun}. The large-distance asymptotics of the RGSF of
$c$-type is \cite{ronveaux1995heun}%
\begin{align}
&  \Pi^{\left(  3\right)  }\left(  \eta,r\right)  \overset{r\rightarrow
\infty}{\sim}\frac{1}{r}\exp\left(  i\left[  \eta\left(  r-M\right)  +\left(
2\eta M+M\frac{\mu^{2}}{\eta}\right)  \ln\left(  \frac{r}{M}-1\right)
+\chi\left(  \eta,M\right)  \right]  \right)  ,\label{crgsf1}\\
&  \Pi^{\left(  4\right)  }\left(  \eta,r\right)  \overset{r\rightarrow
\infty}{\sim}\frac{1}{r}\exp\left(  -i\left[  \eta\left(  r-M\right)  +\left(
2\eta M+M\frac{\mu^{2}}{\eta}\right)  \ln\left(  \frac{r}{M}-1\right)
+\chi\left(  \eta,M\right)  \right]  \right)  , \label{crgsf2}%
\end{align}
where $\chi\left(  \eta,M\right)  $ is the phase of the RGSF of $c$-type
\cite{ronveaux1995heun}. The asymptotics at $r\rightarrow\infty$ of the two
solutions, (\ref{crgsf1}) and (\ref{crgsf2}), satisfies the boundary condition
(\ref{phiinfinite}). For high-energy scattering, we have
\cite{ronveaux1995heun}%
\begin{equation}
\chi\left(  \eta,M\right)  =-2M\eta\ln2-\frac{l\pi}{2}-2M\eta\pi+O\left(
\frac{1}{M\eta}\right)  . \label{Chi}%
\end{equation}

\subsubsection{Exact solution of outgoing wave function: Outgoing
Eddington-Finkelstein coordinate \label{outEdd}}

In order to solve the scattering wave function, we use the
Eddington-Finkelstein coordinate.

\textit{Scalar equation with outgoing Eddington-Finkelstein coordinate.} To
solve the outgoing wave function, we use the outgoing Eddington-Finkelstein
coordinate $(u,r)$ with $u=t-r_{\ast}$ \cite{carroll2004spacetime}. By the
outgoing Eddington-Finkelstein coordinate, we have
\begin{equation}
ds^{2}=-\left(  1-\frac{2M}{r}\right)  du^{2}-2dudr+r^{2}d\theta^{2}+r^{2}%
\sin^{2}\theta d\phi^{2}.
\end{equation}

It is worthy to note that to calculate the Hawking radiation, one needs to
know the discontinuous jump of the outgoing wave function on the horizon. The
outgoing wave function can be solved under the outgoing Eddington-Finkelstein
coordinate and the discontinuous jump can be obtained directly from the
outgoing wave function without the analytic extension. This is because the
outgoing Eddington-Finkelstein coordinate does not diverge at $r=2M$ and,
then, holds for both regions inner and outer the horizon.

The scalar equation (\ref{KG}) then becomes%
\begin{equation}
\left[  -\frac{1}{r^{2}}\frac{\partial}{\partial u}r^{2}\frac{\partial
}{\partial r}-\frac{1}{r^{2}}\frac{\partial}{\partial r}r^{2}\frac{\partial
}{\partial u}+\frac{1}{r^{2}}\frac{\partial}{\partial r}r^{2}\left(
1-\frac{2M}{r}\right)  \frac{\partial}{\partial r}+\frac{L^{2}}{r^{2}}-\mu
^{2}\right]  \Phi=0.
\end{equation}
By means of the variable separation%
\begin{equation}
\Phi=\beta\left(  u,r\right)  Y_{lm}\left(  \theta,\phi\right)  ,
\end{equation}
we arrive at%
\begin{equation}
\left[  -\frac{1}{r^{2}}\frac{\partial}{\partial u}r^{2}\frac{\partial
}{\partial r}-\frac{1}{r^{2}}\frac{\partial}{\partial r}r^{2}\frac{\partial
}{\partial u}+\frac{1}{r^{2}}\frac{\partial}{\partial r}r^{2}\left(
1-\frac{2M}{r}\right)  \frac{\partial}{\partial r}-\frac{l\left(  l+1\right)
}{r^{2}}-\mu^{2}\right]  \beta\left(  u,r\right)  =0. \label{outgoingeq}%
\end{equation}
This equation can be solved exactly.

\textit{Outer horizon.} Outer the horizon, the outgoing solution of Eq.
(\ref{outgoingeq}) reads \cite{ronveaux1995heun}%

\begin{align}
\beta^{outer}\left(  u,r\right)   &  =e^{-i\sqrt{\eta^{2}+\mu^{2}}u}%
e^{-i\sqrt{\eta^{2}+\mu^{2}}r}e^{i\eta r}e^{i2M\eta\left(  1-r/M\right)
}\nonumber\\
&  \times\operatorname*{Hc}{}^{\left(  a\right)  }\left(  -iM\eta,i2M\left(
\eta+\sqrt{\eta^{2}+\mu^{2}}\right)  +iM\frac{\mu^{2}}{\eta}+1,1,1+i4M\sqrt
{\eta^{2}+\mu^{2}},\sigma^{outer},1-\frac{r}{2M}\right)  , \label{Betaout}%
\end{align}
where $\operatorname*{Hc}{}^{\left(  a\right)  }\left(  \alpha,\beta
,\gamma,\delta,\sigma,z\right)  $ is the confluent Heun function and%
\begin{equation}
\sigma^{outer}=l\left(  l+1\right)  +4M^{2}\left(  \eta^{2}+\frac{\mu^{2}}%
{2}\right)  -i2M\left(  \eta+\sqrt{\eta^{2}+\mu^{2}}\right)  .
\label{sigmaout}%
\end{equation}

\textit{Inner horizon. }Inner the horizon, the outgoing solution of Eq.
(\ref{outgoingeq}) reads \cite{ronveaux1995heun}%
\begin{align}
\beta^{inner}\left(  u,r\right)   &  =e^{-i\sqrt{\eta^{2}+\mu^{2}}u}%
e^{-i\sqrt{\eta^{2}+\mu^{2}}r}e^{i\eta r}\left(  r-2M\right)  ^{-i4\sqrt
{\eta^{2}+\mu^{2}}M}e^{i2M\eta\left(  1-r/M\right)  }\nonumber\\
&  \times\operatorname*{Hc}{}^{\left(  a\right)  }\left(  -iM\eta,i2M\left(
\eta-\sqrt{\eta^{2}+\mu^{2}}\right)  +iM\frac{\mu^{2}}{\eta}+1,1,1-i4M\sqrt
{\eta^{2}+\mu^{2}},\sigma^{inner},1-\frac{r}{2M}\right)  , \label{Betain}%
\end{align}
where%
\begin{equation}
\sigma^{inner}=l\left(  l+1\right)  +4M^{2}\left(  \eta^{2}+\frac{\mu^{2}}%
{2}\right)  -i2M\left(  \eta-\sqrt{\eta^{2}+\mu^{2}}\right)  . \label{sigmain}%
\end{equation}

It can be seen that the solutions (\ref{Betaout}) and (\ref{Betain}) satisfy
the boundary conditions given in section \ref{boundarycondition}:
$\beta^{outer}\left(  u,r\right)  \sim e^{-i\omega t}e^{i\omega r_{\ast}}$ and
$\beta^{inner}\left(  u,r\right)  \sim e^{-i\omega t}e^{i\omega r_{\ast}}$
when $r\rightarrow2M$.

\textit{Scattering wave function. }For scattering, we only concern ourselves
with the wave function outside the horizon, so the exact irregular solution
reads%
\begin{equation}
e^{-i\omega t}\Pi^{\left(  3\right)  }\left(  \eta,r\right)  =\beta
^{outer}\left(  u,r\right)  .
\end{equation}
Here we have directly written out the time-dependent part.

\subsubsection{Exact solution of ingoing wave function: Ingoing
Eddington-Finkelstein coordinate}

\textit{Scalar equation with ingoing Eddington-Finkelstein coordinate. }For
the ingoing wave function, instead of the outgoing Eddington-Finkelstein
coordinate, we use the ingoing Eddington-Finkelstein coordinate $(v,r)$ with
$v=t+r_{\ast}$ \cite{carroll2004spacetime}. By the ingoing
Eddington-Finkelstein coordinate, we have%

\begin{equation}
ds^{2}=-\left(  1-\frac{2M}{r}\right)  dv^{2}+2dvdr+r^{2}d\theta^{2}+r^{2}%
\sin^{2}\theta d\phi^{2}.
\end{equation}

Similarly, the ingoing wave function can be solved under the ingoing
Eddington-Finkelstein coordinate and the discontinuous jump can be obtained
directly from the ingoing wave function without the analytic extension. This
is because the ingoing Eddington-Finkelstein coordinate does not diverge at
$r=2M$ and, then, holds for both regions inner and outer the horizon.

Eq. (\ref{KG}) then becomes%
\begin{equation}
\left[  \frac{1}{r^{2}}\frac{\partial}{\partial v}r^{2}\frac{\partial
}{\partial r}+\frac{1}{r^{2}}\frac{\partial}{\partial r}r^{2}\frac{\partial
}{\partial v}+\frac{1}{r^{2}}\frac{\partial}{\partial r}r^{2}\left(
1-\frac{2M}{r}\right)  \frac{\partial}{\partial r}-\frac{L^{2}}{r^{2}}-\mu
^{2}\right]  \Phi=0
\end{equation}
By means of the variable separation%
\begin{equation}
\Phi=\alpha\left(  v,r\right)  Y_{lm}\left(  \theta,\phi\right)  ,
\end{equation}
we arrive at%
\begin{equation}
\left[  \frac{1}{r^{2}}\frac{\partial}{\partial v}r^{2}\frac{\partial
}{\partial r}+\frac{1}{r^{2}}\frac{\partial}{\partial r}r^{2}\frac{\partial
}{\partial v}+\frac{1}{r^{2}}\frac{\partial}{\partial r}r^{2}\left(
1-\frac{2M}{r}\right)  \frac{\partial}{\partial r}-\frac{l\left(  l+1\right)
}{r^{2}}-\mu^{2}\right]  \alpha\left(  v,r\right)  =0. \label{ingoing}%
\end{equation}

\textit{Outer horizon. }Outer the horizon, the ingoing solution of Eq.
(\ref{ingoing}) reads \cite{ronveaux1995heun}%
\begin{align}
\alpha^{outer}\left(  v,r\right)   &  =e^{-i\sqrt{\eta^{2}+\mu^{2}}v}%
e^{i\sqrt{\eta^{2}+\mu^{2}}r}e^{i\eta r}\left(  2M-r\right)  ^{4i\sqrt
{\eta^{2}+\mu^{2}}M}e^{2iM\eta\left(  1-r/M\right)  }\nonumber\\
&  \times\operatorname*{Hc}{}^{\left(  a\right)  }\left(  -iM\eta,i2M\left(
\eta+\sqrt{\eta^{2}+\mu^{2}}\right)  +iM\frac{\mu^{2}}{\eta}+1,1,1+i4M\sqrt
{\eta^{2}+\mu^{2}},\sigma^{outer},1-\frac{r}{2M}\right)  , \label{alphaout}%
\end{align}
where $\sigma^{outer}$ is given by Eq. (\ref{sigmaout}).

\textit{Inner horizon. }Inner the horizon, the ingoing solution of Eq.
(\ref{ingoing}) reads \cite{ronveaux1995heun}
\begin{align}
\alpha^{inner}\left(  v,r\right)   &  =e^{-i\sqrt{\eta^{2}+\mu^{2}}v}%
e^{i\sqrt{\eta^{2}+\mu^{2}}r}e^{i\eta r}e^{i2M\eta\left(  1-\frac{r}%
{M}\right)  }\nonumber\\
&  \times\operatorname*{Hc}{}^{\left(  a\right)  }\left(  -iM\eta,i2M\left(
\eta-\sqrt{\eta^{2}+\mu^{2}}\right)  +iM\frac{\mu^{2}}{\eta}+1,1,1-i4M\sqrt
{\eta^{2}+\mu^{2}},\sigma^{inner},1-\frac{r}{2M}\right)  , \label{alphain}%
\end{align}
where $\sigma^{inner}$ is given by Eq. (\ref{sigmain}).

$\alpha^{inner}\left(  v,r\right)  $ is an analytic extension of
$\alpha^{outer}\left(  v,r\right)  $. It can be seen that the solutions Eqs.
(\ref{alphaout}) and (\ref{alphain}) satisfy the boundary conditions given in
section \ref{boundarycondition}: $\alpha^{outer}\left(  v,r\right)  \sim
e^{-i\omega t}e^{-i\omega r_{\ast}}$ and $\alpha^{inner}\left(  v,r\right)
\sim e^{-i\omega t}e^{-i\omega r_{\ast}}$ when $r\rightarrow2M$.

\textit{Scattering wave function. }For scattering, we only concern ourselves
with the wave function outside the horizon, so the exact irregular solution
reads%
\begin{equation}
e^{-i\omega t}\Pi^{\left(  4\right)  }\left(  \eta,r\right)  =\alpha
^{outer}\left(  v,r\right)  .
\end{equation}

\subsection{Scattering phase shift}

\subsubsection{Formal expression}

To calculate the scattering phase shift, we first express the regular solution
$\Pi\left(  \eta,r\right)  $ which satisfies the boundary condition at $r=2M$
as a linear combination of the irregular solutions $\Pi^{\left(  3\right)
}\left(  \eta,r\right)  $ and $\Pi^{\left(  4\right)  }\left(  \eta,r\right)
$ which satisfy the boundary condition at $r\rightarrow\infty$:%
\begin{equation}
\Pi\left(  \eta,r\right)  =C_{l}i^{l+1}\Pi^{\left(  4\right)  }\left(
\eta,r\right)  +D_{l}\left(  -i\right)  ^{l+1}\Pi^{\left(  3\right)  }\left(
\eta,r\right)  . \label{XiPi1Pi2}%
\end{equation}
Here, the coefficients $C_{l}$ and $D_{l}$ is determined by
\cite{joachain1975quantum}%
\begin{align}
C_{l}  &  =\left(  -i\right)  ^{l+1}\frac{W_{r}\left[  \Pi^{\left(  3\right)
}\left(  \eta,r\right)  ,\Pi\left(  \eta,r\right)  \right]  }{W_{r}\left[
\Pi^{\left(  3\right)  }\left(  \eta,r\right)  ,\Pi^{\left(  4\right)
}\left(  \eta,r\right)  \right]  },\label{cl}\\
D_{l}  &  =-i^{l+1}\frac{W_{r}\left[  \Pi^{\left(  4\right)  }\left(
\eta,r\right)  ,\Pi\left(  \eta,r\right)  \right]  }{W_{r}\left[  \Pi^{\left(
3\right)  }\left(  \eta,r\right)  ,\Pi^{\left(  4\right)  }\left(
\eta,r\right)  \right]  }, \label{dl}%
\end{align}
where the Wronskian determinant $W_{x}\left[  f\left(  x\right)  ,g\left(
x\right)  \right]  =f\left(  x\right)  g^{\prime}\left(  x\right)  -f^{\prime
}\left(  x\right)  g\left(  x\right)  $.

The $S$-matrix can be obtained by Eqs. (\ref{cl}) and (\ref{dl}):%
\begin{equation}
S_{l}\left(  \eta\right)  =e^{2i\delta_{l}}=\frac{D_{l}}{C_{l}}=-\left(
-1\right)  ^{l+1}\frac{W_{r}\left[  \Pi^{\left(  4\right)  }\left(
\eta,r\right)  ,\Pi\left(  \eta,r\right)  \right]  }{W_{r}\left[  \Pi^{\left(
3\right)  }\left(  \eta,r\right)  ,\Pi\left(  \eta,r\right)  \right]  }.
\label{Sl}%
\end{equation}
The scattering phase shift then reads%
\begin{equation}
\delta_{l}\left(  \eta\right)  =\frac{1}{2}\arg\left(  -\left(  -1\right)
^{l+1}\frac{W_{r}\left[  \Pi^{\left(  4\right)  }\left(  \eta,r\right)
,\Pi\left(  \eta,r\right)  \right]  }{W_{r}\left[  \Pi^{\left(  3\right)
}\left(  \eta,r\right)  ,\Pi\left(  \eta,r\right)  \right]  }\right)  .
\label{sps}%
\end{equation}
The scattering phase shift (\ref{sps}) can be rewritten as%
\begin{equation}
\delta_{l}\left(  \eta\right)  =W_{r}\left[  \Pi^{\left(  4\right)  }\left(
\eta,r\right)  ,\Pi\left(  \eta,r\right)  \right]  +\frac{l\pi}{2},
\label{phaseshift}%
\end{equation}
since $W_{r}\left[  \Pi^{\left(  3\right)  }\left(  \eta,r\right)  ,\Pi\left(
k,r\right)  \right]  $ and $W_{r}\left[  \Pi^{\left(  4\right)  }\left(
\eta,r\right)  ,\Pi\left(  \eta,r\right)  \right]  $ are complex conjugate to
each other.

In principle, one can calculate the phase shift by Eq. (\ref{phaseshift}) with
the asymmetrics of the exact wave function, Eqs. (\ref{regularsolution}),
(\ref{Betaout}), and (\ref{alphaout}) directly. The expression Eq.
(\ref{sps}), however, is too complicated, so we provide an explicit
approximation under the weak-field in the next section.

\subsubsection{Weak-field approximation}

In this section, we calculate the scattering phase shift under the weak-field approximation.

In section \ref{scatbohanshu}, we take $p=-ic$ and\ $\beta=i\zeta$, so $c$ and
$\zeta$ are real numbers: $c=M\eta$ and $\zeta=\frac{M}{\eta}\left(  2\eta
^{2}+\mu^{2}\right)  $.\ \ \ \ 

In weak-field approximation, i.e., $c\ll1$ ($M\eta\ll1$), the confluent Heun
equation (\ref{eqyz}) reduces to%
\begin{equation}
\frac{d}{dz}\left(  z^{2}-1\right)  \frac{d}{dz}y\left(  z\right)  +\left[
c^{2}\left(  z^{2}-1\right)  +2c\zeta z-\lambda\right]  y\left(  z\right)  =0.
\label{eqyz1}%
\end{equation}
The asymptotic solution at $z\rightarrow\infty$ ($r\rightarrow\infty$) of Eq.
(\ref{eqyz1}) reads \cite{abramov1979phaseshifts}%
\begin{equation}
\Pi\left(  c,\zeta\right)  \sim\frac{1}{z}\sin\left(  cz-\zeta\ln
2cz-\frac{l\pi}{2}+\delta_{l}\right)  .
\end{equation}
For $c\ll1$ ($M\eta\ll1$), the $S$-matrix can be approximately expressed as
\cite{abramov1979phaseshifts}%
\begin{equation}
e^{2i\delta_{l}}\sim\frac{\Gamma\left(  l+1-i\zeta\right)  }{\Gamma\left(
l+1+i\zeta\right)  }\exp\left(  \frac{2c^{2}\zeta}{\left(  2l-1\right)
\left(  2l+3\right)  }\right)  \frac{1-i\rho e^{il\pi}\frac{\Gamma\left(
l+1+i\zeta\right)  }{\Gamma\left(  -l+i\zeta\right)  }}{1+i\rho e^{-il\pi
}\frac{\Gamma\left(  l+1-i\zeta\right)  }{\Gamma\left(  -l-i\zeta\right)  }},
\end{equation}
where
\begin{equation}
\rho\sim\left(  \frac{c}{4}\right)  ^{2l+1}\frac{\pi^{2}}{\left(  l+\frac
{1}{2}\right)  ^{2}\Gamma^{4}\left(  l+\frac{1}{2}\right)  }\left[
1-c^{2}\frac{4\left(  2l+1\right)  }{\left(  2l-1\right)  ^{2}\left(
2l+3\right)  ^{2}}\left(  \zeta^{2}+\frac{1}{4}\right)  \right]  .
\end{equation}

Then we arrive at
\begin{align}
\Pi\left(  \eta,r\right)   &  \sim\frac{1}{\frac{r}{M}-1}\sin\left(
M\eta\left(  \frac{r}{M}-1\right)  -\frac{M}{\eta}\left(  2\eta^{2}+\mu
^{2}\right)  \ln2M\eta\left(  \frac{r}{M}-1\right)  -\frac{l\pi}{2}+\delta
_{l}\right) \nonumber\\
&  \sim\frac{1}{r}\sin\left(  \eta\left(  r-M\right)  -\frac{M}{\eta}\left(
2\eta^{2}+\mu^{2}\right)  \ln\left(  \frac{r}{M}-1\right)  -\frac{l\pi}%
{2}+\delta_{l}\right)  .
\end{align}
and
\begin{align}
e^{2i\delta_{l}}  &  \sim\frac{\Gamma\left(  l+1-i\left[  \frac{M}{\eta
}\left(  2\eta^{2}+\mu^{2}\right)  \right]  \right)  }{\Gamma\left(
l+1+i\left[  \frac{M}{\eta}\left(  2\eta^{2}+\mu^{2}\right)  \right]  \right)
}\exp\left(  2M^{2}\eta^{2}\left[  \frac{M}{\eta}\left(  2\eta^{2}+\mu
^{2}\right)  \right]  \frac{1}{\left(  2l-1\right)  \left(  2l+3\right)
}\right) \nonumber\\
&  \times\frac{1-i\rho e^{il\pi}\frac{\Gamma\left(  l+1+i\left[  \frac{M}%
{\eta}\left(  2\eta^{2}+\mu^{2}\right)  \right]  \right)  }{\Gamma\left(
-l+i\left[  \frac{M}{\eta}\left(  2\eta^{2}+\mu^{2}\right)  \right]  \right)
}}{1+i\rho e^{-il\pi}\frac{\Gamma\left(  l+1-i\left[  \frac{M}{\eta}\left(
2\eta^{2}+\mu^{2}\right)  \right]  \right)  }{\Gamma\left(  -l-i\left[
\frac{M}{\eta}\left(  2\eta^{2}+\mu^{2}\right)  \right]  \right)  }}.
\end{align}

When the mass $M$ is very small, we have
\begin{equation}
e^{2i\delta_{l}}\sim\frac{\Gamma\left(  l+1-i2\eta M\right)  }{\Gamma\left(
l+1+i2\eta M\right)  }.
\end{equation}
This is just the scattering phase shift of the Newtonian inverse-square potential.

\subsection{Asymptotic behavior of scattering wave function}

In scattering, we often concern the large-distance asymptotic behavior of the
scattering wave function. The asymptotics at $r\rightarrow\infty$ of the
scattering wave function (\ref{XiPi1Pi2}) can be achieved in virtue of Eqs.
(\ref{crgsf1}) and (\ref{crgsf2}):%

\begin{equation}
\Pi\left(  \eta,r\right)  \overset{r\rightarrow\infty}{\sim}\frac{1}{r}%
\sin\left(  \eta r+\left(  2\eta M+M\frac{\mu^{2}}{\eta}\right)  \ln\left(
\frac{r}{M}-1\right)  +\delta_{l}+\chi\left(  \eta,M\right)  -\frac{l\pi}%
{2}-\eta M\right)  . \label{xi}%
\end{equation}
The asymptotic expression of $\chi\left(  \eta,M\right)  $ for high-energy
scattering is given by Eq. (\ref{Chi}).

\section{Bound-state eigenvalue: Exact solution \label{exacteigenvalue}}

Now we return to the problem of bound states. Based on the solution of
scattering obtained above, we present the eigenvalue of the bound state.

In section \ref{Eigenvalue}, we provide an explicit asymptotic expression for
the eigenvalue of bound states. Here we provide an exact implicit expression
for the eigenvalue of bound states.

The eigenvalue spectrum of bound states, according to the $S$-matrix theory,
is the singularity of the $S$-matrix on the positive real axis
\cite{joachain1975quantum}. Taking $\eta=ik$, we have%
\begin{equation}
S_{l}\left(  k\right)  =-\left(  -1\right)  ^{l+1}\frac{W_{r}\left[
\Pi^{\left(  4\right)  }\left(  k,r\right)  ,\Pi\left(  k,r\right)  \right]
}{W_{r}\left[  \Pi^{\left(  3\right)  }\left(  k,r\right)  ,\Pi\left(
k,r\right)  \right]  }. \label{slk}%
\end{equation}
The singularity of the $S$-matrix (\ref{slk}) is just the zero of the
denominator \cite{joachain1975quantum} $W_{r}\left[  \Pi^{\left(  3\right)
}\left(  k,r\right)  ,\Pi\left(  k,r\right)  \right]  $ with $k>0$, i.e.,%
\begin{equation}
W_{r}\left[  \Pi^{\left(  3\right)  }\left(  k,r\right)  ,\Pi\left(
k,r\right)  \right]  =0,\text{ }k>0.\text{\ } \label{Wr}%
\end{equation}
This is an exact implicit expression of the eigenvalue of bound states. The
bound-state eigenvalue $-k^{2}$ can be solved from Eq. (\ref{Wr}).

\section{Discontinuous jump on horizon: Jump condition \label{jumpcondition}}

The scattering wave function on the horizon $r=2M$ has a discontinuous jump.
In order to calculate the scattering wave function, we need to first determine
the jump of the wave function on the horizon, i.e., the jump condition.

In literature, the discontinuous jump of the wave function is obtained by
analytically extending an asymptotic solution
\cite{hawking1975particle,damour1976black}.

In this paper, without analytic extension, we calculate the discontinuous jump
from the exact result obtained in the present paper directly. The reason why
we need not to perform the analytic extension is that the coordinate adopted
here is the Eddington-Finkelstein coordinate which has no singularity on the
horizon and holds for both regions inner and outer the horizon.

\subsection{Jump condition: Outgoing wave function}

In section \ref{outEdd}, we obtain the exact solution of the outgoing wave
function outside the horizon, Eq. (\ref{Betain}), and obtain the outgoing wave
function inside the horizon, Eq. (\ref{Betaout}), under the
Eddington-Finkelstein coordinate. Comparing the wave functions inside and
outside the horizon, we can directly obtain the jump of the wave function on
the horizon.

By the outgoing solutions (\ref{Betaout}) and (\ref{Betain}), the jump of the
outgoing wave function on the horizon $r=2M$ is
\begin{equation}
\left.  \frac{\beta^{outer}\left(  u,r\right)  }{\beta^{inner}\left(
u,r\right)  }\right\vert _{r\rightarrow2M}=\lim_{r\rightarrow2M^{-}}\frac
{1}{\left(  r-2M\right)  ^{-i4\sqrt{\eta^{2}+\mu^{2}}M}}=e^{-4M\sqrt{\eta
^{2}+\mu^{2}}\pi}, \label{Edd1}%
\end{equation}
where $\operatorname*{Hc}{}^{\left(  a\right)  }\left(  \alpha,\beta
,\gamma,\delta,\sigma,0\right)  =1$ is used in the calculation. Note that the
limit here is a left limit since the factor $\left(  r-2M\right)
^{-i4\sqrt{\eta^{2}+\mu^{2}}M}$ comes from $\beta^{inner}\left(  u,r\right)  $.

It can be directly seen that there exists a jump on the phase of the
scattering wave function.

In literature, e.g., Refs. \cite{hawking1975particle,damour1976black}, the
jump is calculated from a asymptotic solution. Our result agrees the result in literature.

\subsection{Jump condition: Ingoing wave function}

Similarly, by the exact ingoing scattering wave functions inside and outside
the horizon obtained in section \ref{outEdd}, we can directly calculate the
jump on the horizon of the ingoing wave function.

By the ingoing solutions (\ref{alphaout}) and (\ref{alphain}), the jump of the
ingoing wave function on the horizon $r=2M$ is%
\begin{equation}
\left.  \frac{\alpha^{outer}\left(  v,r\right)  }{\alpha^{inner}\left(
v,r\right)  }\right\vert _{r\rightarrow2M}=\lim_{r\rightarrow2M^{+}}\left(
2M-r\right)  ^{4i\sqrt{\eta^{2}+\mu^{2}}M}=e^{-4M\sqrt{\eta^{2}+\mu^{2}}\pi}.
\label{jumpingoing}%
\end{equation}

Comparing Eqs. (\ref{Edd1}) and (\ref{jumpingoing}) shows that the jump of the
outgoing wave and the jump of the ingoing wave are the same. Note that the
limit here is a right limit since the factor $\left(  2M-r\right)
^{4i\sqrt{\eta^{2}+\mu^{2}}M}$ comes from $\alpha^{outer}\left(  v,r\right)  $.

\subsection{Discontinuous jump on horizon: Comparison with Hawking and
Damour-Ruffini methods \label{comparison}}

In order to calculate the Hawking radiation, one needs to know the
discontinuous jump of the wave function on the horizon.

Hawking and Ruffini dealt with this problem from different viewpoints
respectively. They, mathematically speaking, did almost the same thing. The
methods they used are both based on asymptotic solutions.

In the above, we have obtained the exact wave functions inner and outer the
horizon, so we can calculate the jump exactly rather than asymptotically.

It should be emphasized that the starting point of all the methods is the
solution of the outgoing wave of the radial equation (\ref{radialeq}).

In the following, we compare our method with the Hawking treatment and the
Damour and Ruffini treatment.

\subsubsection{Hawking method: Brief review}

Hawking \cite{hawking1975particle} and Damour and Ruffini
\cite{damour1976black} both start from the $r\rightarrow\infty$ ($r_{\ast
}\rightarrow\infty$) asymptotics of the outgoing wave function%
\begin{equation}
\beta^{outer}\overset{r\rightarrow\infty}{\sim}\frac{1}{r}e^{-i\omega u}%
=\frac{1}{r}e^{2i\omega r_{\ast}}e^{-i\omega v}, \label{rout}%
\end{equation}
where $u$ and $v$ are the outgoing and ingoing Eddington-Finkelstein
coordinates, respectively.

Hawking in Ref. \cite{hawking1975particle} uses quantum field theory to deal
with the radiation of a black hole.

In Hawking's treatment, the outgoing wave function $\Phi_{\omega}%
^{outer}=\beta^{outer}Y_{lm}\left(  \theta,\phi\right)  $ is expanded as%
\begin{equation}
\Phi_{\omega}^{outer}=\int d\omega^{\prime}\left(  a_{\omega\omega^{\prime}%
}f_{\omega^{\prime}}+b_{\omega\omega^{\prime}}\bar{f}_{\omega^{\prime}%
}\right)
\end{equation}
with the expansion coefficients%
\begin{align}
a_{\omega\omega^{\prime}}  &  \sim\frac{1}{2\pi}\left(  \frac{\omega^{\prime}%
}{\omega}\right)  \left(  -\omega^{\prime}\right)  ^{-i\frac{\omega}{\kappa
}-1},\label{acoe}\\
b_{\omega\omega^{\prime}}  &  =-i\alpha_{\omega\left(  -\omega^{\prime
}\right)  }, \label{bcoe}%
\end{align}
where $f_{\omega}$ and $\bar{f}_{\omega}$ are the solutions on past null
infinity $f^{-}\left(  t=-\infty,r=+\infty\right)  $ containing only positive
frequencies and only negative frequencies, respectively. The number of
particles created by the gravitational field and emitted to infinity then
reads \cite{hawking1975particle}
\begin{equation}
\left\langle 0_{-}\right\vert N_{\omega}\left\vert 0_{-}\right\rangle =\int
d\omega^{\prime}\left\vert b_{\omega\omega^{\prime}}\right\vert ^{2}.
\label{Nomiga}%
\end{equation}
Notice that the subscript of $\alpha_{\omega\left(  -\omega^{\prime}\right)
}$ in Eq. (\ref{bcoe}) is $-\omega^{\prime}$. In order to calculate the
integral in Eq. (\ref{Nomiga}), we need to analytic extend $\omega^{\prime}$.
In Eq. (\ref{acoe})\ $\omega^{\prime}=0$ is a singular point. Hawking in Ref.
\cite{hawking1975particle} performs the analytic extension by anticlockwisely
rounding this singularity%
\begin{equation}
\omega^{\prime}\rightarrow\omega^{\prime}e^{-i\pi}. \label{hawking}%
\end{equation}
By virtue of such an analytic extension, Eq. (\ref{bcoe}) becomes%
\begin{equation}
a_{\omega\left(  -\omega^{\prime}\right)  }=ie^{-\omega/\kappa}a_{\omega
\omega^{\prime}}.
\end{equation}
That is to say, $a_{\omega\omega^{\prime}}$ has a discontinuous jump at
$\omega=0$%
\begin{equation}
\frac{a_{\omega\left(  -\omega^{\prime}\right)  }}{a_{\omega\omega^{\prime}}%
}=ie^{-\omega/\kappa}.
\end{equation}
Accordingly, Hawking arrives at%
\begin{equation}
\left\langle 0_{-}\right\vert N_{\omega}\left\vert 0_{-}\right\rangle
=\frac{1}{e^{\frac{\omega}{kT}}-1},\text{\quad\quad}T=\frac{1}{8\pi Mk}.
\end{equation}

\subsubsection{Damour-Ruffini method: Brief review}

Damour and Ruffini, also starting from the asymptotic outgoing wave
function\ (\ref{rout}), deferent from Hawking, use scattering method to deal
with the radiation of a black hole.

Damour and Ruffini rewrite the asymptotic outgoing wave outside the horizon
(\ref{rout}) as \cite{damour1976black}%
\begin{equation}
\beta^{outer}\overset{r\rightarrow2M}{\sim}\frac{1}{r}e^{2i\omega r_{\ast}%
}e^{-i\omega v}=\frac{1}{r}e^{2i\omega r}e^{-i\omega v}\left(  \frac{r-2M}%
{2M}\right)  ^{4i\omega M}\nonumber
\end{equation}
to expose the singularity at the horizon $r=2M$. In order to analytically
extend the outgoing wave to the inner horizon, they analytically extend $r-2M$
as \cite{damour1976black}%
\begin{align}
r-2M  &  \rightarrow\left\vert r-2M\right\vert e^{-i\pi}\nonumber\\
&  =\left(  2M-r\right)  e^{-i\pi}. \label{ruffini}%
\end{align}
The outgoing wave in the horizon then reads%
\begin{equation}
\beta^{inner}=\frac{1}{r}e^{4\pi\omega M}e^{2i\omega r_{\ast}}e^{-i\omega v}.
\end{equation}
There\ is also a discontinuous jump%
\begin{equation}
\frac{\beta^{outer}}{\beta^{inner}}=e^{-4\pi\omega M}. \label{betafac}%
\end{equation}
Then they obtain a relative scattering probability \cite{damour1976black}%
\begin{equation}
P_{\omega}=e^{-8\pi\omega M}%
\end{equation}
and then obtain the Hawking's result:%
\begin{equation}
N_{\omega}=\frac{1}{e^{8\pi\omega M}-1}=\frac{1}{e^{\frac{\omega}{kT}}%
-1},\text{ \ }T=\frac{1}{8\pi Mk}.
\end{equation}

\subsubsection{Direct calculation through exact solution: Comparison}

The starting points of Hawking method and Damour-Ruffini method are both the
$r\rightarrow2M$ asymptotic outgoing wave function (\ref{rout}), though their
treatments are somewhat different.

In this paper, using the outgoing Eddington-Finkelstein coordinate, we obtain
the exact solution of\ the outgoing wave function outside the horizon, Eq.
(\ref{Betaout}), and we also obtain the exact result inside the horizon, Eq.
(\ref{Betain}) by analytic extension. This allows us to calculate the jump exactly.

From the exact results, we can calculate the jump (\ref{Edd1}) exactly:
\begin{equation}
\left.  \frac{\beta^{outer}}{\beta^{inner}}\right\vert _{r\rightarrow
2M}=e^{-4M\sqrt{\eta^{2}+\mu^{2}}\pi}=e^{-4M\omega\pi}.
\end{equation}

The result obtained in the present is based on the exact solution rather than
the asymptotic solution.

\section{Conclusion \label{conclusion}}

An exact solution for a scalar field in the Schwarzschild spacetime is
provided. For bound states, we obtain an exact bound-state wave function and
an exact implicit expression of eigenvalues. For scattering, we obtain an
exact scattering wave function.

Moreover, besides the exact solutions, we also provide some approximate
solutions. For bound states, we provide an explicit asymptotic expression for
bound-state eigenvalues. For scattering, and an approximate expression under
the weak-field approximation for phase shift.

It is worthy to note that the solution of the scalar equation inner the
horizon can be obtained by analytically extending the outer solution. One may
also discuss the solution inner the horizon further based on interior metrics
of a Schwarzschild spacetime \cite{doran2008interior,wenda1985globally}.

In literature, the discontinuous jump on horizon of the wave function is
obtained by analytically extending an asymptotic solution
\cite{hawking1975particle,damour1976black}. In this paper, since we have
obtained the exact solution of wave functions, we can calculate this
discontinuous jump straightforwardly from the exact result without analytic extensions.

Scattering by a Schwarzschild spacetime is a long-range potential scattering.
In further researches, we can attempt the heat-kernel method to calculate the
scattering phase shift using the approach developed in Refs.
\cite{graham2009spectral,barvinsky1987beyond,barvinsky1990covariant,barvinsky1990covariant3,mukhanov2007introduction,pang2012relation,li2015heat}%
. Moreover, starting from the exact solution, we can investigate the
large-distance asymptotic behavior which is important in the scattering
problem
\cite{koyama2001asymptotic,hod2013scattering,liu2014scattering,li2016scattering}%
. Starting from the exact solution given in the present paper, we can also
calculate the corresponding heat kernel and the partition function
\cite{vassilevich2003heat}. Using the heat kernel, we can calculate various
quantum field theory quantities, such as one-loop effective actions, vacuum
energies, etc. \cite{dai2009number,dai2010approach}.

\acknowledgments

We are very indebted to Dr G. Zeitrauman for his encouragement. This work is supported in part by NSF of China under Grant
No. 11575125 and No. 11675119. We are grateful to Prof. Ming-Ming Kang of Sichuan University for pointing out the problem of the sign in Eqs.  (4.12), (4.13), (4.15), and (4.16).





\end{document}